\documentclass[]{article}

\usepackage{pifont}
\usepackage{array}
\usepackage{booktabs}

\usepackage{balance}
\usepackage{hyperref}

\usepackage{amsmath}
\usepackage{enumitem}
\usepackage{xspace}

\usepackage{url}
\usepackage{tablefootnote}

\usepackage{listings}

\usepackage[table]{xcolor}

\usepackage[capitalize,noabbrev]{cleveref}

\usepackage{algorithmic}
\usepackage{graphicx}
\usepackage{textcomp}
\usepackage{booktabs}
\usepackage{verbatim}
\usepackage{balance}
\usepackage{hyperref}
\usepackage{array}

\title{Characterizing Data Scientists in the Real World}

\author{Paula Pereira$^{1}$ \and Jácome Cunha$^{1,2}$ \and João P. Fernandes$^{3}$\\
a77672@alunos.uminho.pt, jacome@fe.up.pt, jpf@dei.uc.pt\\
$^{1}$University of Minho\\
$^{2}$HASLab/INESC TEC\\
$^{3}$CISUC, University of Coimbra\\
Portugal}

\date{}

\begin{document}

\maketitle

\begin{abstract}

Data collection is pervasively bound to our digital lifestyle. A recent study by the IDC reports that the growth of the data created and replicated in 2020 was even higher than in the previous years due to pandemic-related confinements to an astonishing global amount of 64.2 zettabytes of data. 

While not all the produced data is meant to be analyzed, there are numerous companies whose services/products rely heavily on data analysis. That is to say that mining the produced data has already revealed great value for businesses in different sectors. But to be able to fully realize this value, companies need to be able to hire professionals that are capable of gleaning insights and extracting value from the available data.

We hypothesize that people nowadays conducting data-science-related tasks in practice may not have adequate training or formation. So in order to be able to fully support them in being productive in their duties, e.g. by building appropriate tools that increase their productivity, we first need to characterize the current generation of data scientists. To contribute towards this characterization, we conducted  a public survey to fully understand \textit{who} is doing data science, \textit{how} they work, what are the \textit{skills} they hold and lack, and which \textit{tools} they use and need. 

\end{abstract}

\noindent
\textbf{Keywords -- } data science, survey, empirical evidence

	\section{Introduction}\label{sec:introduction}
    Every day huge amounts of data are created and mined to glean insights and produce value~\cite{Holst2019,Kubina2015,Parkins2017}. According to an estimation from DOMO, every minute 6 million people shop online, 5.7 million searches are conducted on Google and there are 575 thousand posts on Twitter.\footnote{https://www.domo.com/learn/infographic/data-never-sleeps-9} We observe similar trends in other domains such as aviation: Forbes magazine reported that, for every flight, five to eight terabytes of data are generated regarding e.g., the passengers, the condition of the engines; this corresponds to 30 times more data than those generated by the previous generation aircraft~\cite{Maire2016}.

Companies from all sectors have realized the value of their data and are trying to use it to gain competitive advantage~\cite{Davenport2012, Grupe1995, Kubina2015, Provost2013, Williams2018}. The interest in exploring the potential of the data that is collected is reflected in its appreciation, making it one of the most valuable resources to organizations. In fact, in May 2017 \textit{The Economist} stated that oil was no longer the world's most valuable resource, losing that position to data~\cite{Parkins2017} and that \textit{data scientist} has even been classified as the \textit{sexiest job of the 21\textsuperscript{st} century}~\cite{Davenport2012}.


Due to the large variety and high volume of available data, as well as the various data analytics solutions, companies are looking to have in their teams people with great skills in gathering, cleaning, and using data~\cite{Davenport2012, Dhar2014, Provost2013}.  The challenge is that it has been found that the number of job offers in this area exceeds demand~\cite{Perry2019}. As a consequence, companies are presumably hiring workers to conduct data science tasks even if their background and/or formation does not necessarily match the ideal profile. This impacts the heterogeneity in their data science workflow which remains unknown and understudied and consequently makes the development of supporting methodologies and tools more challenging. 

Although many professionals have been working on data analysis and mining for much longer, it is only a little over a decade since ‘data scientist’ has been recognized as a professional occupation~\cite{Cao2017}. Moreover, only recently, researchers have started studying and understanding this community~\cite{Wongsuphasawat2019}.

To fully understand how we can assist data science workers in being more productive in their jobs, we first need to understand who they are, how they work, what skills they hold and lack, and which tools they need. Given their importance in practice, the data science community should focus on these people, see how they usually work and ask them which challenges they face. 

The main goal of our work is thus to answer the following research questions (RQs):

\begin{itemize}
    \item[$\mathcal{RQ}_1$] What is the profile of a data science professional?
    \item[$\mathcal{RQ}_2$] How does the profile of the data science professional impact their work?
    \item[$\mathcal{RQ}_3$] Which technologies are most frequently used?
\end{itemize}


We believe this knowledge will be highly beneficial to the data science community as well as those who create new data science tools, as they will be able to deliver  software that is more reliable and accessible to the various groups of employees they target.

To accomplish this goal we conducted a public survey distributed to data science workers. 








	\section{Related Work}
	
One may find in the literature studies that are related to ours.

Schroeck et al. realized a study consisting of a survey of 1144 business and IT professionals across 95 countries and interviews with over two dozen academics, subject matter experts, and business executives~\cite{Schroeck2012}. The study aimed to understand if the interviewees were adopting big data techniques to their advantage. It concluded that, compared to previous years, there had been significant growth in respondents whose companies practiced tasks on top of large amounts of data and that this data was of great importance to their organizations. It was also possible to understand that, contrary to popular belief, big data was perceived by professionals as more than just a huge volume of social media data. Although this work followed a similar methodology to our own, it focused on the advantages that organizations had when using large amounts of data, not taking into consideration who are the individuals that use that data, how they work with that data and what are the challenges they face when preparing and analyzing it.



Rexer et al. conducted a survey to better understand the behavior, views, and preferences of data analytics professionals~\cite{Rexer2015}. It consisted of 59 questions which were e-mailed to over 10000 data science professionals. It provided a more complete report than~\cite{Schroeck2012} focusing on the professionals, the tasks they were responsible for, and the tools and programming languages that they used. It is interesting to note that in this survey, and compared to previous years, more professionals described themselves as data scientists.


Mack and Tarrant realized an even more complete study, complementing a survey with in-depth interviews with 19 high-level managers and professionals on how they approach data science skills in their organizations \cite{Mack2015}. This study showed that the demand for data scientists remained quite strong, with data collection, data analysis, data interpretation, and visualization skills being the most desired capabilities. Despite its completeness, this study was limited to only two groups of professionals (data scientists and managers).



Harris et al. described the results of a survey on data scientists, their experiences and how they viewed their own skills and careers \cite{Harris2013}. This survey was designed by data scientists themselves, who used its results to identify a new and more precise vocabulary for talking about data science work, based on how data scientists describe themselves and their skills. Through the results they showed that tools are critical to data scientists’ effectiveness. Although they managed to distinguish several sub-groups of professionals, this information could have been enriched if they had taken into account the academic training of the participants, and also their preferences regarding the tools and techniques they use.

Miryung et al.  conducted 16 interviews with data scientists from eight different product organizations within Microsoft to understand their responsibilities, considering their education and training backgrounds, their missions in software engineering contexts, and the type of problems in which they worked \cite{Kim2016}. The authors then characterized the roles of data scientists at Microsoft, having identified five different styles of work (insight providers, modeling specialists, platform builders, polymaths, and team leaders). However, the results might not be representative considering all the interviewees worked at Microsoft.

 Muller et al.  conducted several interviews with 21 data science professionals \cite{Muller2019}. These interviews allowed the authors to focus on the way data scientists work with their data. The authors found that they are involved in various steps of the process and perform tasks like data collection, data cleaning, data integration, and feature engineering. They also showed that, often, data is not ready for analysis, and must be designed to meet the requirements of an algorithm. 

Zhang et al. focused on the collaboration of data workers during the several steps of a data science workflow  \cite{Zhang2020}. To do so, they conducted an online survey with 183 participants who work in various aspects of data science and learned that data science teams are extremely collaborative and work with a variety of stakeholders and tools during a data science project. Similarly to the previously mentioned study conducted by Miryung et al. \cite{Kim2016}, the results of both of these studies may not be representative considering that all the respondents worked in the same company (IBM).

    \section{Survey design and analysis methodology}
    \label{survey}
    Because we wanted to study the impact of academic background and professional experience in the way professionals of data science work, we designed a survey to collect information from professionals around the world. 
We now present information on the survey design and distribution (\cref{sec:design}). Furthermore, we present how the collected data was prepared for analysis (\cref{sec:prep}) and the analysis methodology used (\cref{sec:anal}). 

\subsection{Survey design and distribution}\label{sec:design}

Bearing in mind that the quality of the results of a survey is greatly affected by the quality of the questions that compose it, during its preparation, care was taken to write the questions to ensure that they were not ambiguous or too complex to answer \cite{Bickman2008, Wohlin2012}. 
The survey was divided into six sections as follows:

\begin{enumerate}
    \item Academic background; 
    \item Professional situation; 
    \item Self-evaluation of strengths on several tasks;
    \item Work characterization (problems, time spent coding);
    \item Technologies used;
    \item Demographic questions.
\end{enumerate}

The survey was developed on Google Forms, and was distributed online to data science workers via forums (Stack Overflow, Kaggle, Reddit, Facebook, LinkedIn) and email to people we knew were working in the field or could distribute the survey among such people. To enhance data collection, the survey accepted answers from April of 2020 to the end of the same year. 

\subsection{Data preparation}\label{sec:prep}

 Data cleaning deals with removing anomalies from a dataset in order to ensure its quality~\cite{Chu2016,Kandel2011,Muller,Rahm2014}.
Detecting and repairing data with errors is one of the major challenges in data analysis, and failing to do so can result in unreliable decisions and poor study results \cite{Chu2016,VanDenBroeck2005}. For this reason, before attempting to analyze the data collected through the survey, we searched for errors and inconsistencies. In our case, two types of anomalies could potentially affect data quality: missing data and contradictions. 
    

Concerning missing values, all fields were mandatory, the exceptions being fields related to the academic background, since the person may not have any formal training, and also the fields related to the difficulties experienced in the performed tasks, assuming that, in cases of lack of response, the person is not affected by such situation. Regarding contradictions, we detected some cases in which the number of degrees indicated by the person was incompatible with the indicated age (e.g. a participant indicated 13 different degrees). In these cases, and because our analysis relies on the information about the participant's academic background, the responses were not considered for analysis.

    
Additionally, we filtered the attributes that are relevant to the analysis, as well as derived new fields \cite{Kitchenham2003}. Therefore, the dataset was also analyzed to check whether all the information captured was still relevant for the intended analysis. We proceeded to filter the relevant attributes and created a new one called \textit{CS background?} which, taking into account the information related to academic background, was populated with ``Y''  when the person indicated any formal training in Computer Science, and with ``N'' otherwise.
    
At the end of the data preparation process, we had 116 responses and 35 attributes containing information regarding the academic background of the participants, their work, the difficulties they face, and the technologies they use.

\subsection{Analysis methodology}\label{sec:anal}
\label{subsec:analysis_methodology}

To analyze the data we applied two techniques: descriptive statistics and inferential statistics, which we detail next.

\paragraph{Descriptive statistics} These
depend on the level of measurement, a mechanism by which a variable is scored. Generally, there are three broad levels of measurement: nominal, ordinal, and continuous~\cite{Fisher2009}.
The results of the survey are nominal, meaning that there is no hierarchy between the categories from which the participants chose their answers. To analyze this type of data, we used two types of measurement. As a measurement of central tendency, we calculated the mode for each question, i.e. the value with the greatest frequency. As a measurement of dispersion, we determined the frequency distribution, i.e. the number of cases of each category.  

Because one of the goals of the survey is to understand if there are differences between people with different backgrounds and experiences, we applied multivariate descriptive statistics to explore whether there are relations between more than one variable. To represent this information, we used contingency tables, where each cell represents the intersection of two variables of interest.

With descriptive statistics, we were able to draw conclusions about the surveyed participants. However, we also considered this information useful to make reasonable affirmations about the larger population. To do so, we used inferential statistics.

\paragraph{Inferential statistics}

These are calculated with the purpose of generalizing the findings from a sample to the entire population of interest~\cite{Shane2009}. The tests applied in inferential statistics also depend on the type of data. Because the data obtained is nominal, and some of the frequencies observed are lower than $5$, we used Fisher's exact test to evaluate whether the results from the survey can be generalized to the entire population.


Fisher's test is a technique for hypothesis testing based on data in the form of frequencies when the sample is small. This test is performed to verify if the results obtained in two different groups are statistically significant, to study the relationship between two different nominal variables, and the strength of that relationship~\cite{Shane2009,Kim2017}.
This test was used on the survey's results to evaluate how certain variables, namely Computer Science Background and Years of experience, influenced the others. To apply the test to our dataset, we developed a script in R.
%
For each group being tested, a contingency table is calculated, which is used to compute Fisher's test p-value. 
%
%
The null hypothesis is that there is no difference among the groups being tested. Therefore, whenever p-value is lower than $0.05$, we conclude that there is evidence to reject the null hypothesis, and therefore, that the groups are different.

    \section{Survey Results}
    \label{sec:surv_res}

We now present the results we obtained from our survey.\footnote{For data visualization we used \href{https://powerbi.microsoft.com/}{Microsoft Power BI}}

\subsection{Demographic questions}

We wanted to gather a large number of responses from a diverse group of people in terms of geography, gender, and age. To capture this information, the survey included a section with demographic questions asking participants their gender, their age and their country.
This data is shown in \cref{tab:demographics}.


\begin{table}[!tbh]
\caption{Country demographics.}
\begin{center}
\begin{tabular}{lrrr}
\toprule
\textbf{Country} & \textbf{Female} & \textbf{Male} & \textbf{Total}\\
\midrule
     Australia & 1 & 1 & 2 \\
     Brazil & & 4 & 4 \\
     Canada & 1 & 7 & 8 \\
     Germany & & 6 & 6\\
     India & & 6& 6\\
     Netherlands & 1 & 1 & 2\\
     Norway & 2 & 1 &3\\
     Portugal & 12 & 32 & 44\\
     Spain & 1 & 2 & 3\\
     Switzerland & 1 & 1 & 2\\
     Thailand & 1 & 1 & 2\\
     UK & & 7 & 7\\
     USA & 3 & 9 & 12\\
     Other & 6 & 9 & 15\\
     \midrule
     \textbf{Total} & \textbf{29} & \textbf{87} & \textbf{116}\\
     \bottomrule
\end{tabular}
\label{tab:demographics}
\end{center}
\end{table}

\subsection{Academic background}

To collect information on the academic background of the participants, we asked them to indicate all the degrees (Bachelor, Master, Doctoral, or Professional) and corresponding fields. 

From all participants, 110  (94.8\%) indicated having a bachelor's degree, 90 participants (77.6\%) indicated having a master's degree, 31 participants (26.7\%) a doctoral degree, and 13 participants (11.2\%) received some type of professional training. Only 2 participants (1,7\%) indicated not having any academic degree. This information also showed that 29 participants (25\%) have a bachelor's degree, a master's degree, and a doctoral degree. In general, the most mentioned fields of study were \textit{Computer Science or Computer Engineering or Software engineering}, \textit{Mathematics or Statistics}, and \textit{Another engineering disciplines}. There are also participants with background in \textit{Natural Sciences (biology, chemistry, physics)}, \textit{Social Sciences (anthropology, psychology, political science, sociology)}, and \textit{Humanities (arts, law, languages,  literature)}. Thus, we can infer that, \textbf{despite the heterogeneity of academic backgrounds, data science professionals reported being highly qualified}.

Most participants, 72  (62.07\%), have some formal training in computer science, and 44 participants (37.93\%) have no background in computer science. In the case of female participants, the majority stated that they had no prior computer science experience, with only 41.1\% indicating that they had some training in the field. In the case of male participants, 69\%  indicated having some background in computer science.

\begin{figure}[!tbh]
        \centering
        \includegraphics[width=\columnwidth]{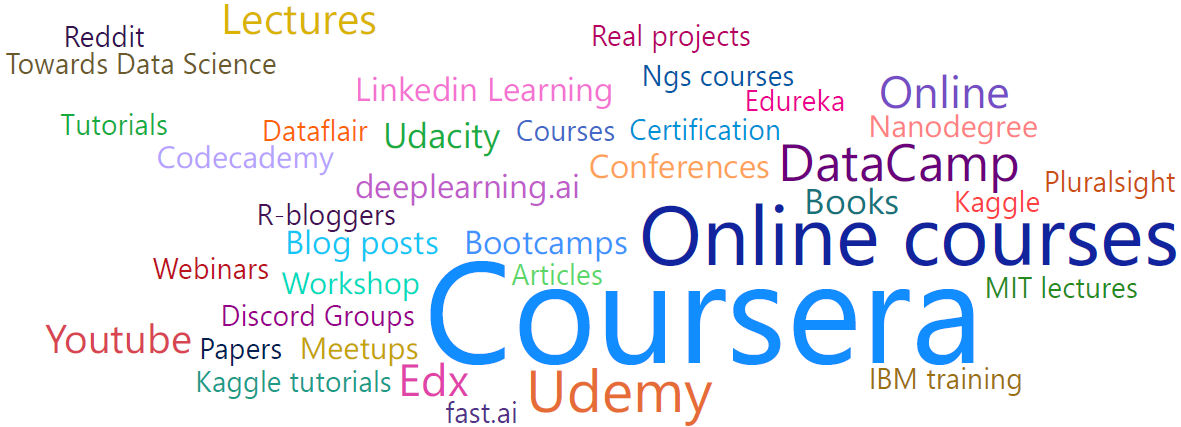}
        \caption{Other learning methods.}
        \label{fig:learning_methods}
    \end{figure}

Besides formal education, the survey included an open-answer question for participants to indicate other methods of learning data science. As shown in \cref{fig:learning_methods}, the platform \textit{Coursera}\footnote{\url{https://www.coursera.org/}}
was mentioned 49 times by the participants. In fact, \textbf{almost every answer to this question mentioned some type of online resource, from online courses to blogs about data science and online lectures from ivy league universities}. Also, as demonstrated in \cref{fig:learning_methods}, several participants referred going to \textit{bootcamps, conferences} and \textit{meetups} as a good learning opportunity, but only 4 participants mentioned \textit{books}. 

\subsection{Professional experience}

To be able to study the professional experience of data science professionals, the survey included a section with questions related to the careers of the participants. The answers in this section reflect the participants' current jobs, the years of experience they have, how satisfied they are with their jobs, the time they spent coding, and their analytical goals.

\paragraph{Current job}
    
Many participants, 51  (44,0\%), indicated working as a \textit{data scientist}, 11 participants (9,5\%) identified their job title as \textit{machine learning engineer}, and another 11 (9,5\%) as \textit{data analysts}. Besides these three job titles, which are some of the most recognized positions in data science, we also received answers from people working as \textit{software developer/engineer} (7,8\%),  \textit{educator or academic researcher} (7,8\%), \textit{consultant} (3,4\%), \textit{database administrator} (0,9\%), \textit{statistician} (0,9\%), \textit{computer scientist} (0,9\%), and \textit{other} (15,5\%). 


\paragraph{Years of professional experience}

Participants were also asked how many years of professional experience they have in data science, and how satisfied they are with their work. 
24 participants (20,69\%) say they have less than 2 years of experience, 48 participants (41.38\%) say they have between 2 and 4 years of experience, 28 participants (24,14\%) say they have between 5 and 9 years of experience, 8 participants (6,90\%) say they have between 10 and 14 years of experience, 6 participants (5,17\%) say they have between 15 and 24 years of experience, and the remaining 2 participants (1,72\%) say they have 25 and 39 years of experience. 

\paragraph{Job satisfaction}
    
Regarding job satisfaction, 
only 1 participant (0,86\%) says he is \textit{extremely dissatisfied}, 11 participants (9,48\%) say they are \textit{slightly dissatisfied}, 20 participants (17,24\%) say they are \textit{neutral}, 41 participants (35,34\%) say they are \textit{slightly satisfied}, and finally, 43 participants (37,07\%) say they are \textit{extremely satisfied} with their jobs. With only 10\% of participants indicating they are somewhat dissatisfied, we can conclude that \textbf{in general, data science workers report being satisfied with their careers.}

To see if the level of satisfaction is influenced by other factors, the results for the question on satisfaction were crossed with the results obtained about the questions on years of experience, background, and gender. For the first two, we could not find significant differences, but for the third we did.
%
%

\begin{figure}[!htb]
        \centering
        \includegraphics[width=1\columnwidth]{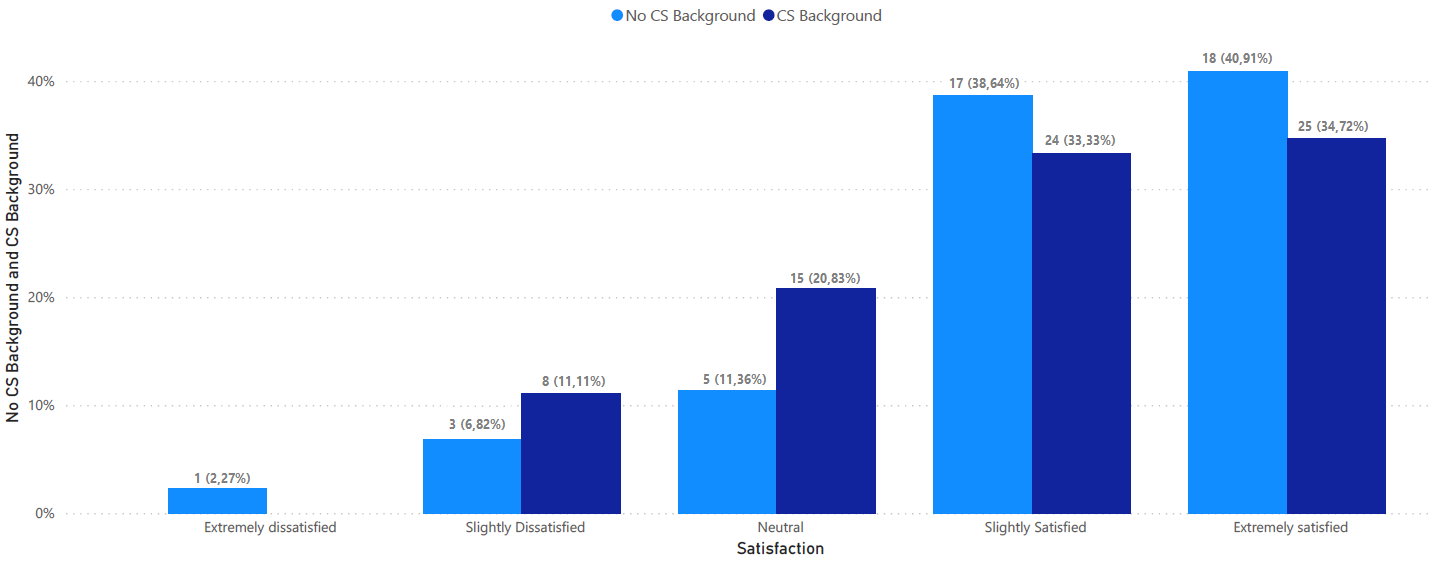}
        \caption{Satisfaction by background.}
        \label{fig:satisfaction_background}
    \end{figure}

In  \cref{fig:satisfaction_background}, the information on satisfaction is  presented taking into account the background in computer science. As can be seen, nearly 80\% of those who did not have a CS background said they are \textit{slightly satisfied}/\textit{extremely satisfied}. This value drops to around 68\% among people with a CS background. Furthermore, the percentage of people with a CS background who are dissatisfied (11.11\%) is also higher than the percentage of people who do not have a CS background and are dissatisfied (9.09\%). \textbf{Thus, the reported satisfaction is lower in people with a CS background.}

\begin{figure}[!htb]
        \centering
        \includegraphics[width=1\columnwidth]{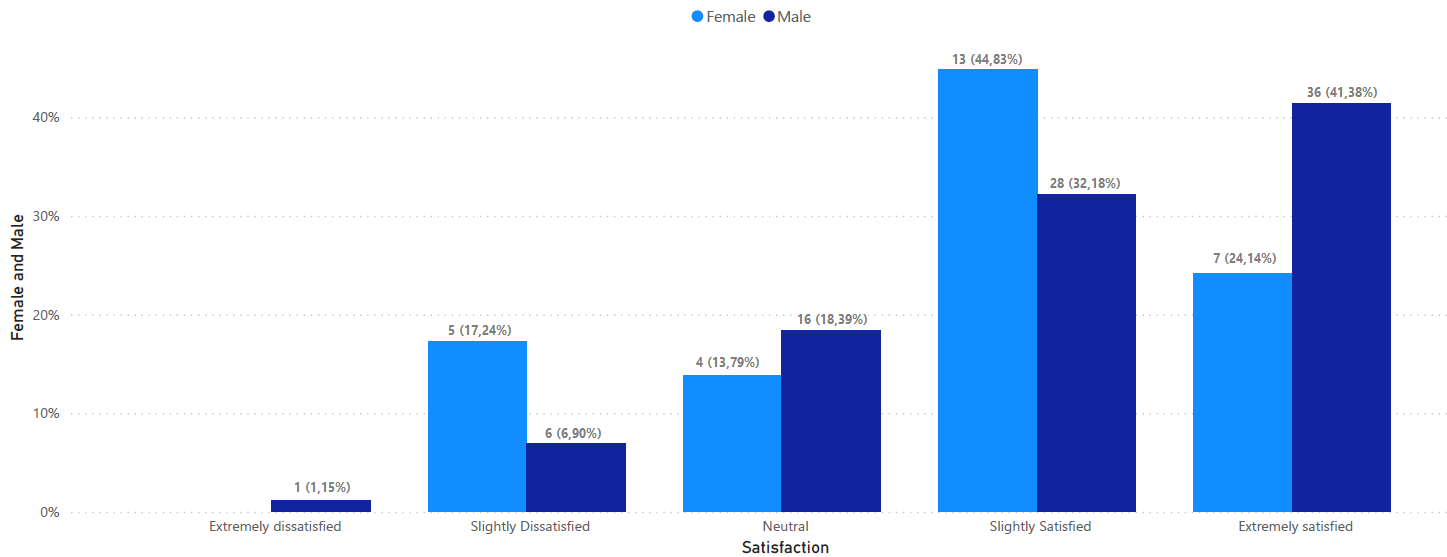}
        \caption{Satisfaction by gender.}
        \label{fig:satisfaction_gender}
    \end{figure}

\cref{fig:satisfaction_gender} shows the satisfaction between female and male participants. Regarding female participants, almost 69\% are \textit{slightly satisfied}/\textit{extremely satisfied}, and 17,24\% are \textit{slightly dissatisfied}. Looking at the results of male participants, while 73,56\%  are \textit{slightly satisfied}/\textit{extremely satisfied}, only 8,05\% says that they are dissatisfied. \textbf{Therefore, satisfaction is lower in women.}

\subsection{Self-evaluation on data science tasks}

To understand how participants assess their competence in tasks related to data manipulation, data processing, and data analysis, they rated their experience in a set of tasks as: \textit{Very Poor} - Little or no knowledge/expertise; \textit{Poor} - Experimental/vague knowledge; \textit{Ok} - Familiar and competent user; \textit{Good} - Regular and confident user; \textit{Very Good} - Leading expert (\cref{table:summary_task}). 


\begin{table}[!bht]
\caption{Strengths in data science tasks (in percentage).}
\label{table:summary_task}
\begin{center}
\begin{tabular}{p{5.4cm}>{\raggedleft\arraybackslash}p{2cm}>{\raggedleft\arraybackslash}p{1cm}>{\raggedleft\arraybackslash}p{2cm}}
\toprule
\multicolumn{1}{c}{\textbf{Task}} & \textbf{Very Poor and Poor} & \textbf{Ok}                     & \textbf{Good and Very Good} \\ \midrule
Translating business problems to data science problems & 4,31                                                         & 28,45                  & 67,24                                                        \\ \midrule
Collecting data & 3,45                                                         & 25                     & 71,55                                                       \\ \midrule
 Assessing the quality of data & 1,72                                                         & 17,24                  & \cellcolor[HTML]{D3D3D3}\textbf{81,04}                                                        \\ \midrule
 Filtering relevant attributes & 1,72                                                         & 19,83                  & \cellcolor[HTML]{D3D3D3}78,45                                                        \\ \midrule
 Extracting new attributes & 3,46                                                         & 31,03                  & 65,51                                                        \\ \midrule
 Cleaning data & 1,72                                                         & 20,69                 & 77,59                                                        \\ \midrule
 Applying data visualization & 11,2                                                         & 25                     & 63,8                                                         \\ \midrule
 Applying classical statistics & \cellcolor[HTML]{D3D3D3}11,21                                                        & 31,9                   & 56,89                                                       \\ \midrule
 Applying data mining techniques & 10,35                                                        & 35,34                  & 54,31                                                        \\ \midrule
 Applying deep learning techniques & \cellcolor[HTML]{D3D3D3}\textbf{30,17}                                                        & 31,9                   & 37,93                                                        \\ \midrule
 Evaluate results to respond to business problems/find opportunities & \cellcolor[HTML]{D3D3D3}11,21                                                        & 24,14                  & 64,65                                                        \\ \midrule
Transmitting acquired knowledge & 3,45                                                         & 18,1                   & \cellcolor[HTML]{D3D3D3}78,45\\
\bottomrule
\end{tabular}
\end{center}
\end{table}

As can be seen in \cref{table:summary_task}, for each task, the percentage of positive responses (\textit{Good} and \textit{Very Good}) is always higher than 50\%,  and the tasks that received a higher percentage of negative responses (\textit{Very Poor} and \textit{Poor}) are tasks related to applying analysis techniques and evaluating the results. These results indicate that, in general, \textbf{professionals assess their performance positively and that they feel competent in all of the tasks described, with data analysis-related tasks being the most challenging.}

\textit{``Assessing the quality of data"} was the task that received the highest percentage of positive responses from participants (81,04\%). During the initial phase of the analysis process, this task is very important, since being able to assess the quality of the data has a significant impact on the rest of the process and the results obtained. \textit{``Transmitting acquired knowledge"} also received a high percentage of positive responses, showing  \textbf{data science professionals, in their majority, report having no trouble in providing meaningful insights that can lead to problem solving.}

Pre-processing related tasks, such as \textit{``Filtering relevant attributes"}, \textit{``Extracting new attributes"} or \textit{``Cleaning data"}, are usually the most time-consuming in data science projects, however, as suggested by the information in \cref{table:summary_task} data science professionals feel capable and comfortable conducting these tasks.

In opposition, \textit{``Applying deep learning techniques"} was the task that received the highest percentage of negative responses from participants (30,17\%). This can be explained considering the higher complexity deep learning techniques pose compared to the traditional machine learning models and statistical analysis. Moreover, there are usually fewer projects where these techniques may apply, due to the use case itself or the number of available resources. This can lead data science professionals to have \textbf{reduced exposure of \textit{deep learning} techniques and, as a consequence, be less familiar, experienced, and comfortable applying them}.

\begin{figure}[!bth]
        \centering
        
        \includegraphics[width=1\columnwidth]{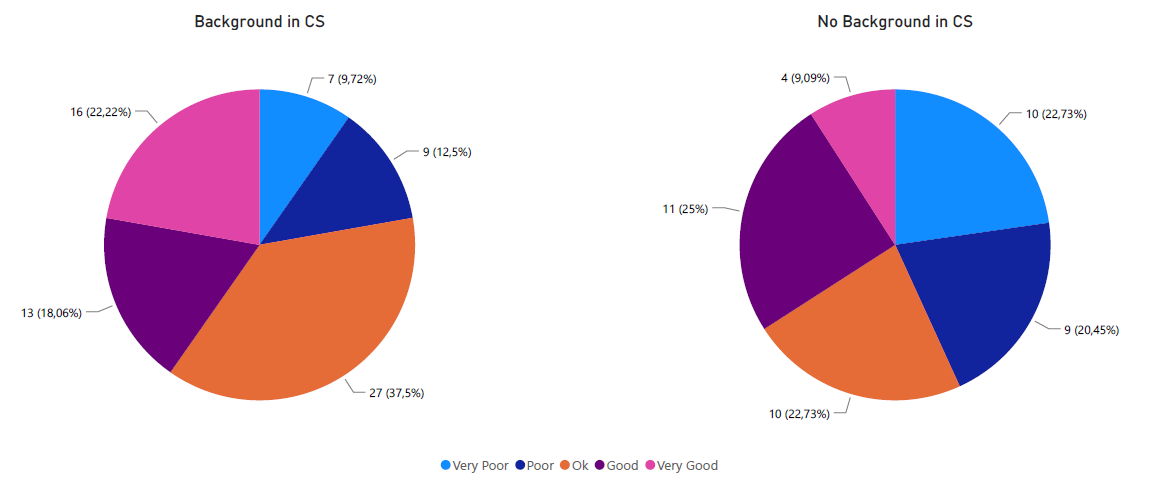}
        \caption{Applying deep learning techniques by Background.}
        \label{fig:deep_back}
    \end{figure}

The perceived experience the respondents reported for this particular task is highly related to whether or not the respondent has a CS background. This was determined by conducting Fischer's test as described in \cref{subsec:analysis_methodology}, which resulted in a $p$-$value$ of $0.049$. Upon further evaluation, we conclude that \textbf{those with a CS background (77,78\%) feel more apt to apply \textit{deep learning} techniques than those without (56,82\%)}, as shown in \cref{fig:deep_back}.

\subsection{Work characterization}

To be able to describe the work of a data science professional, we collected information on the difficulties professionals face daily, on the time they spend coding, and also on their analytical goals.

Regarding difficulties that professionals may face, participants rated how frequently they face some situations as \textit{Never, Rarely, Sometimes, Often}, or \textit{Always}. The results are summarized in \cref{table:summary_prob}.

\begin{table}[!bth]
\caption{Problems that affect data scientists (in percentage).}
\label{table:summary_prob}
\begin{center}
\begin{tabular}{p{5.8cm}>{\raggedleft\arraybackslash}p{2cm}>{\raggedleft\arraybackslash}p{2cm}}
\toprule
\multicolumn{1}{c}{\textbf{Situation description}} & \textbf{Never to Sometimes} & \textbf{Often to Always} \\ \midrule
1. Poor quality data & 33,62 & \cellcolor[HTML]{D3D3D3}\textbf{66,38} \\ \midrule
2. Difficult access to relevant data & 41,38 & \cellcolor[HTML]{D3D3D3}58,62 \\ \midrule
3. Lack of data science skills & \cellcolor[HTML]{D3D3D3}\textbf{83,62} & 16,38 \\ \midrule
4. Lack of clear questions to answer & 50,87 & 49,13 \\ \midrule
5. Lack of domain knowledge & \cellcolor[HTML]{D3D3D3}70,69 & 29,31 \\ \midrule
6. Integrating findings into decisions & 58,63 & 41,37 \\ \midrule
7. Expectations of project impact & 54,31 & 45,69 \\ \midrule
8. Results not used by decision makers & 61,21 & 38,79 \\
\bottomrule
\end{tabular}
\end{center}
\end{table}

As shown, of the various situations listed in \cref{table:summary_prob}, only two received a percentage of \textit{Often to Always} responses higher than 50\%: \textit{``Poor quality data"} and \textit{``Difficult access to relevant data"}. As both  concern data, the main object of study for data science professionals, this shows that \textbf{the fact that there is currently a huge variety of data sources may be making the work of those who use data  more difficult}. 

\begin{figure}[!p]
    \begin{center}
        \centering
        \includegraphics[width=.95\columnwidth]{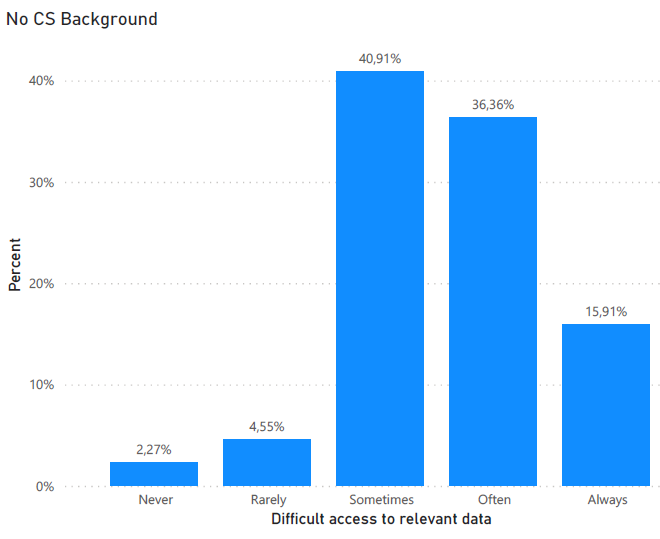}\\
        \includegraphics[width=.95\columnwidth]{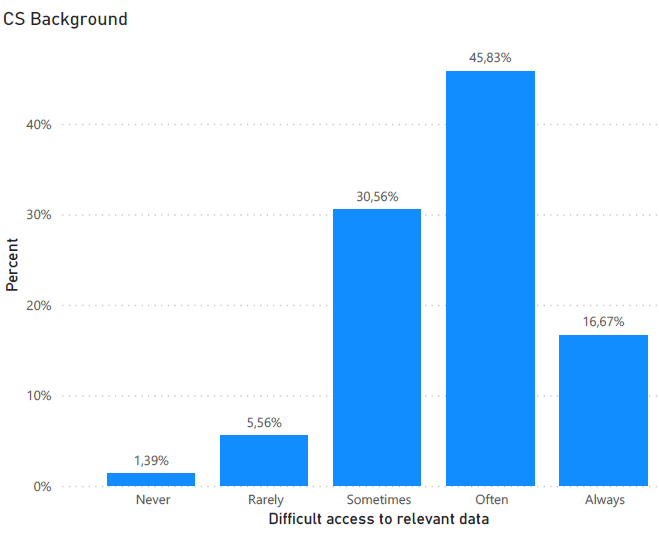}
        \caption{Access to relevant data by background.}
        \label{fig:prob2}
        \end{center}
    \end{figure}

Concerning the \textit{``Difficult access to relevant data''}, we can infer that this situation is justified by the fact that data is being generated at an exponential rate, which can also mean that \textbf{it is necessary to put more thought into the process of data collection}. Besides, as shown in \cref{fig:prob2}, people with a CS background feel more affected by this situation than people without a CS background. 

In addition to these two situations, the one that most frequently affects participants' work is the \textit{``Lack of clear questions to answer"}. 
%
    \begin{figure}[!thbp]
        \centering
        \includegraphics[width=.95\columnwidth]{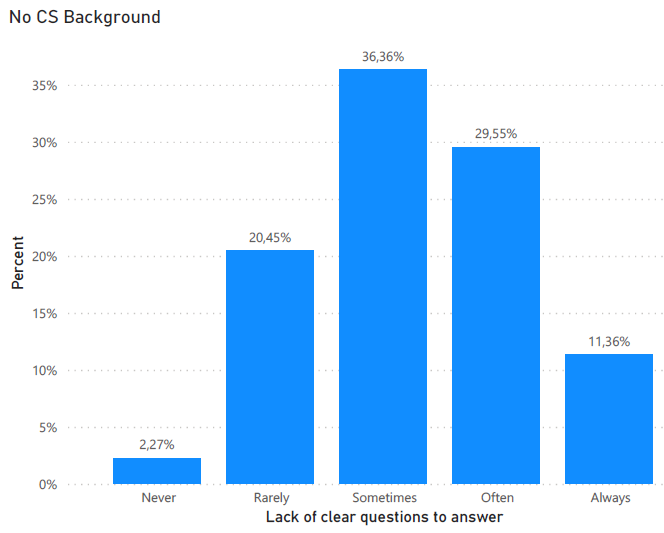}\\
        \includegraphics[width=.95\columnwidth]{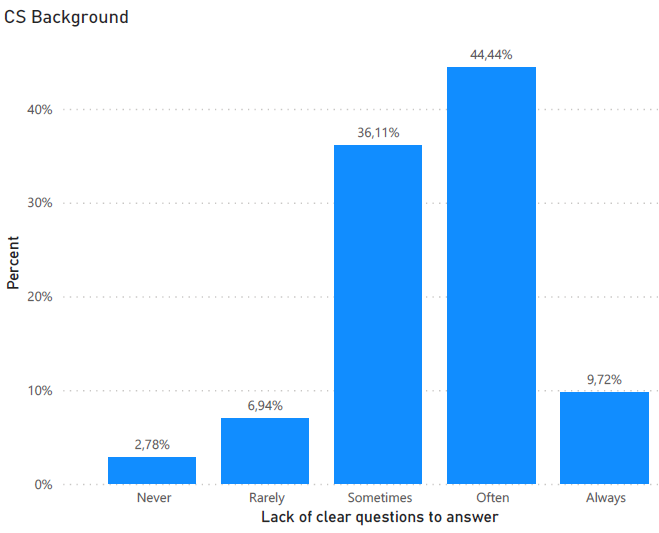}
        \caption{Lack of clear questions to answer by background.}
        \label{fig:prob4}
    \end{figure} 
As shown in \cref{fig:prob4}, \textbf{people with a CS background are more affected by this than people without, which may be justified by the fact that they have a more technological profile, making it more difficult to identify the questions to be answered}.

Analyzing \cref{table:summary_prob}, we also see that the two situations with the highest percentage of \textit{``Never to Sometimes''} responses are: \textit{``Lack of data science skills"} and \textit{``Lack of domain knowledge"}. Because one of the main goals of our work is to understand whether professionals with different backgrounds have a range of different skills that have an impact on the way they do their work, it is worth analyzing the difference in the responses of people with and without a CS background, especially concerning the \textit{``Lack of data science skills"}. 

\begin{figure}[!thbp]
        \centering
        \includegraphics[width=.95\columnwidth]{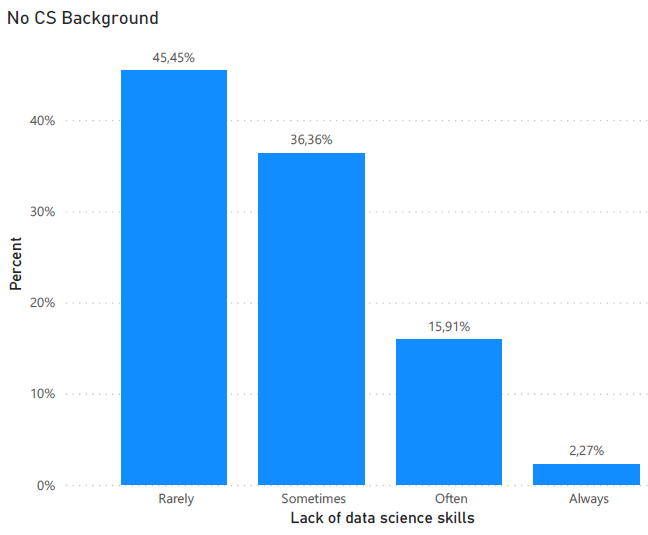}
        \includegraphics[width=.95\columnwidth]{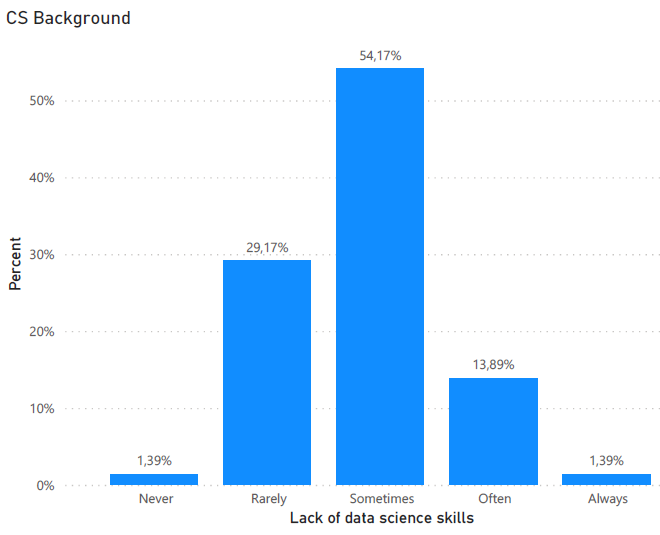}
        \caption{Lack of data science skills by background.}
        \label{fig:prob3}
    \end{figure}

As shown in \cref{fig:prob3}, the biggest difference between the two groups is in the distribution of percentages for \textit{``Rarely"} and \textit{``Sometimes"} responses. As for participants with no CS background, about 45\% of participants answered \textit{``Rarely"}, and about 36\% answered \textit{``Sometimes"}. In the group of participants with a CS background, about 29\% answered \textit{``Rarely"}, and about 54\% answered \textit{``Sometimes"}. Thus, based on the self-reported data, \textbf{the lack of data science skills affects more people who have a CS background}.

\begin{figure}[!htbp]
        \centering
        \includegraphics[width=1\columnwidth]{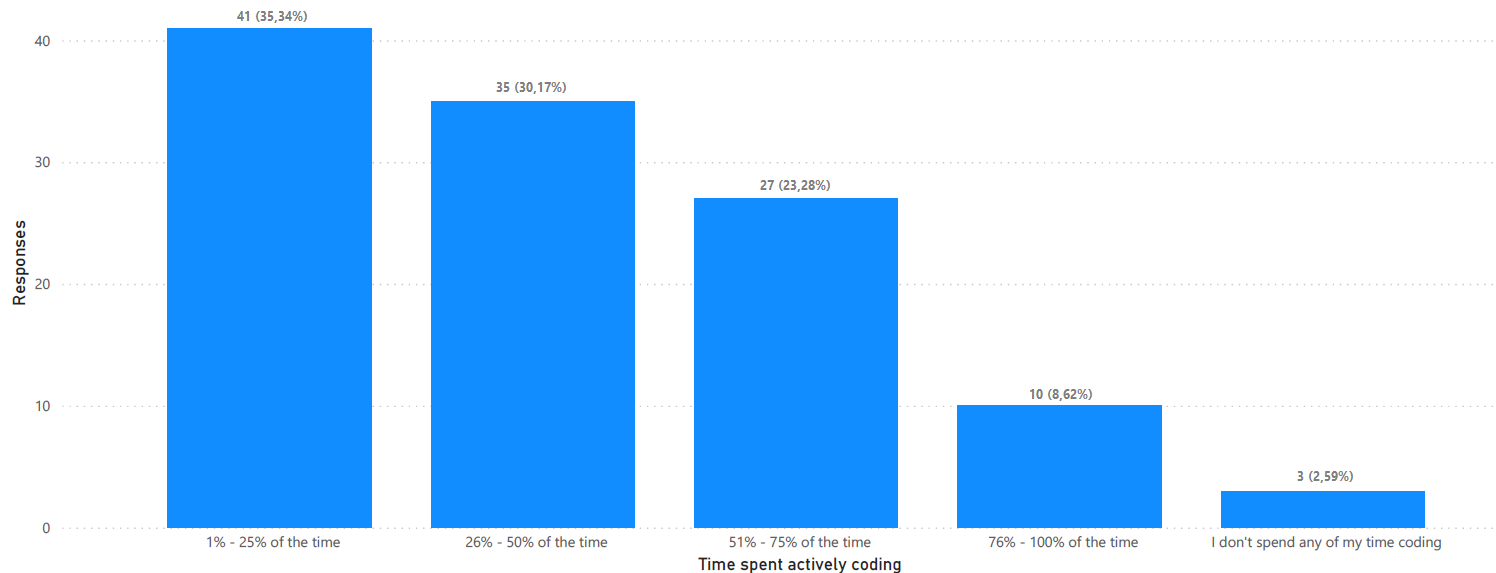}
        \caption{Time spent actively coding.}
        \label{fig:coding}
    \end{figure}

Concerning the time spent actively coding, as shown in \cref{fig:coding}, only 3 participants (2,59\%) say  they \textit{``don't spend any time coding''}. 76 participants, which corresponds to more than 50\%, say they code up to 50\% of their time, with the highest percentage saying they only do it about\textit{``1\% - 25\% of their time''} (35,34\%). Of the remaining participants, 27 people (23,28\%) say they spend \textit{``51\% - 75\% of their time''} coding, and only 10 people (8,62\%) say they spend up to 100\% of their time coding.

\begin{figure}[!htp]
        \centering
        \includegraphics[width=1\columnwidth]{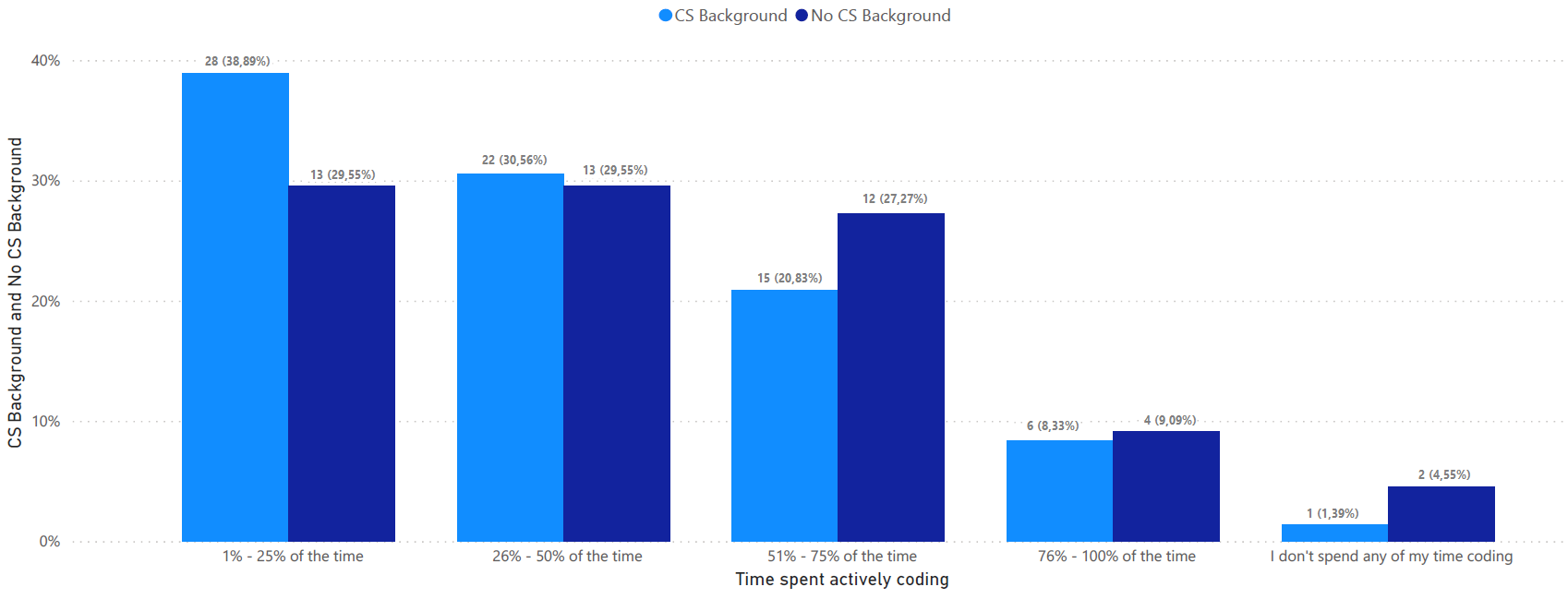}
        \caption{Time spent coding by background.}
        \label{fig:coding_back}
    \end{figure}

In \cref{fig:coding_back}, we can see the difference between people with and without a CS background in the time spent in programming. Concerning people who spend \textit{``1\% - 25\% of their time''} coding, the percentage of people with a CS background is higher (39.89\%) than the percentage of people without a CS background (29.55\%). Also, looking at the bars corresponding to spending more than 50\% of the time coding, we see that the percentage of responses is always higher for people without a CS background. Therefore, we can conclude that \textbf{people without a CS background spend more time coding than people with a CS background.}

\begin{figure}[!htbp]
        \centering
        \includegraphics[width=1\columnwidth]{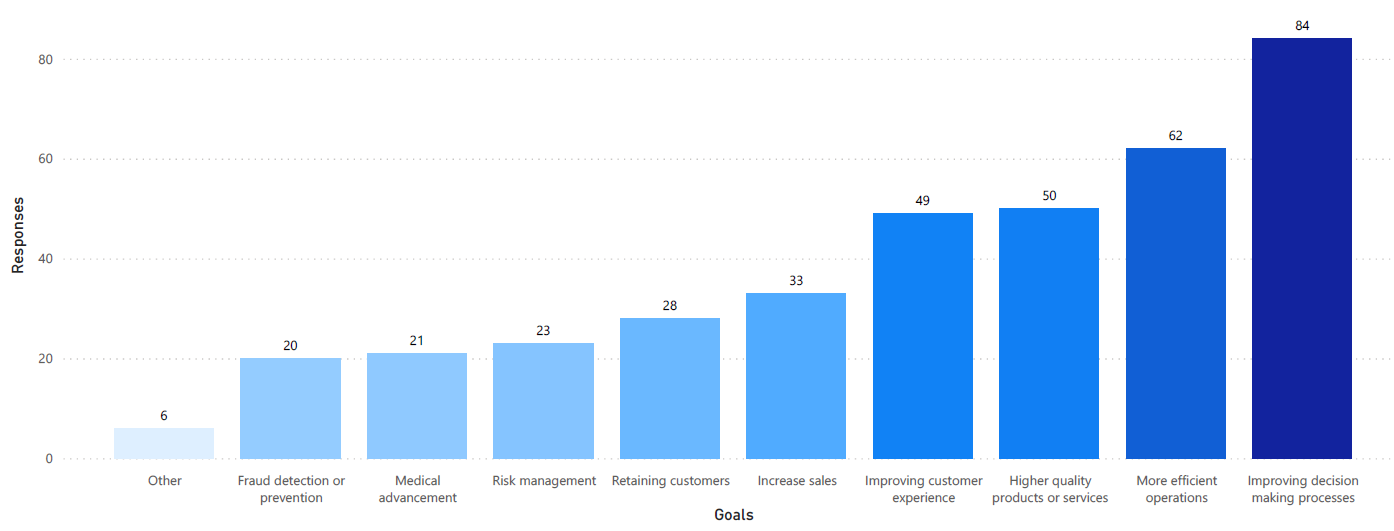}
        \caption{Participants analytical goals.}
        \label{fig:goals}
    \end{figure}

Lastly, the participants indicated their analytical goals (three per participant, on average). As shown in \cref{fig:goals}, the most common goal  is \textit{``Improving decision making processes"}, which was indicated by 84 people (72\%). This is not unexpected since data science has been universally described as a set of principles and techniques that allow decision-making supported by knowledge extracted from data. Next, 62 participants (53\%) say they use data science to achieve \textit{``More efficient operations"}. Less than 20\% of the participants indicated that their analytical goals are \textit{``Fraud detection or prevention"}, \textit{``Medical advancement"}, or \textit{``Risk management"}. 

\subsection{Technology}    

In addition to the analytical component, there is an equally important technological component. To see if the background influences the choices of data science professionals, participants were asked to indicate the technologies they use the most.

\paragraph{IDEs or Editors}

\begin{figure}[!tbhp]
        \centering
        \includegraphics[width=1\columnwidth]{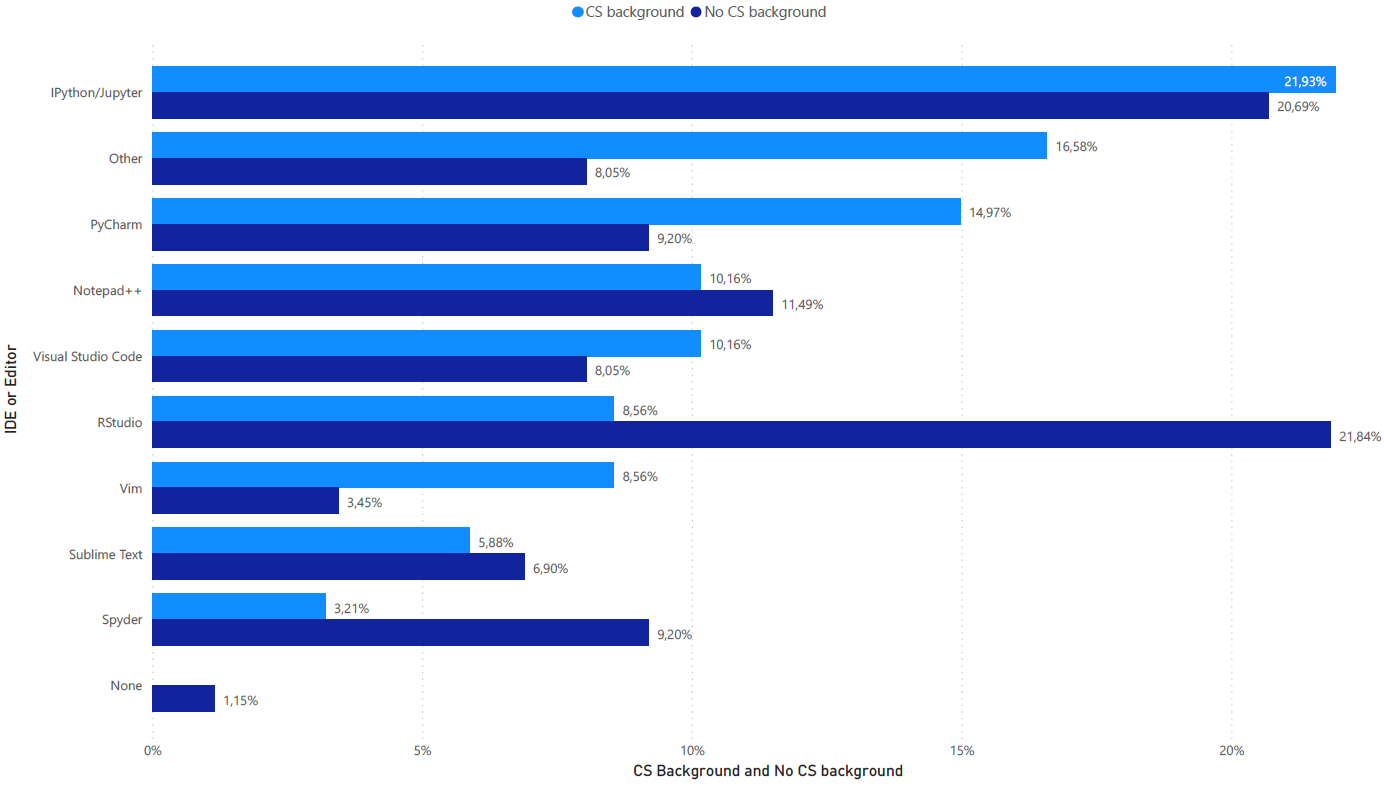}
        \caption{IDE or Editor by background.}
        \label{fig:tech1}
    \end{figure}

\cref{fig:tech1} shows the choices of the participants regarding \textit{IDEs or editors}. The most common IDE  used by people with a CS background is \textit{IPython/Jupyter} (21.93\%). The percentage of use of this editor by people without a CS background is quite similar (20.69\%), making \textit{IPython/Jupyter} the second option with the highest number of responses in this group. The most used editor by people without a CS background is \textit{RStudio}, but its use by people with a CS background is much lower (8.56\%). In the group of people with a CS background, the second most used text editor is \textit{PyCharm} (14.97\%), which reveals a preference for editors for programming in Python.
In addition to the \textit{IDEs} and \textit{text editors}, 16.58\% of people with a CS background indicated using another option. For people without a CS background, this percentage is less than half (8.05\%). This option includes tools such as \textit{IntelliJ} or \textit{Matlab}.

\paragraph{Programming, Scripting or Markup Language}

\begin{figure}[!tbhp]
        \centering
        \includegraphics[width=1\columnwidth]{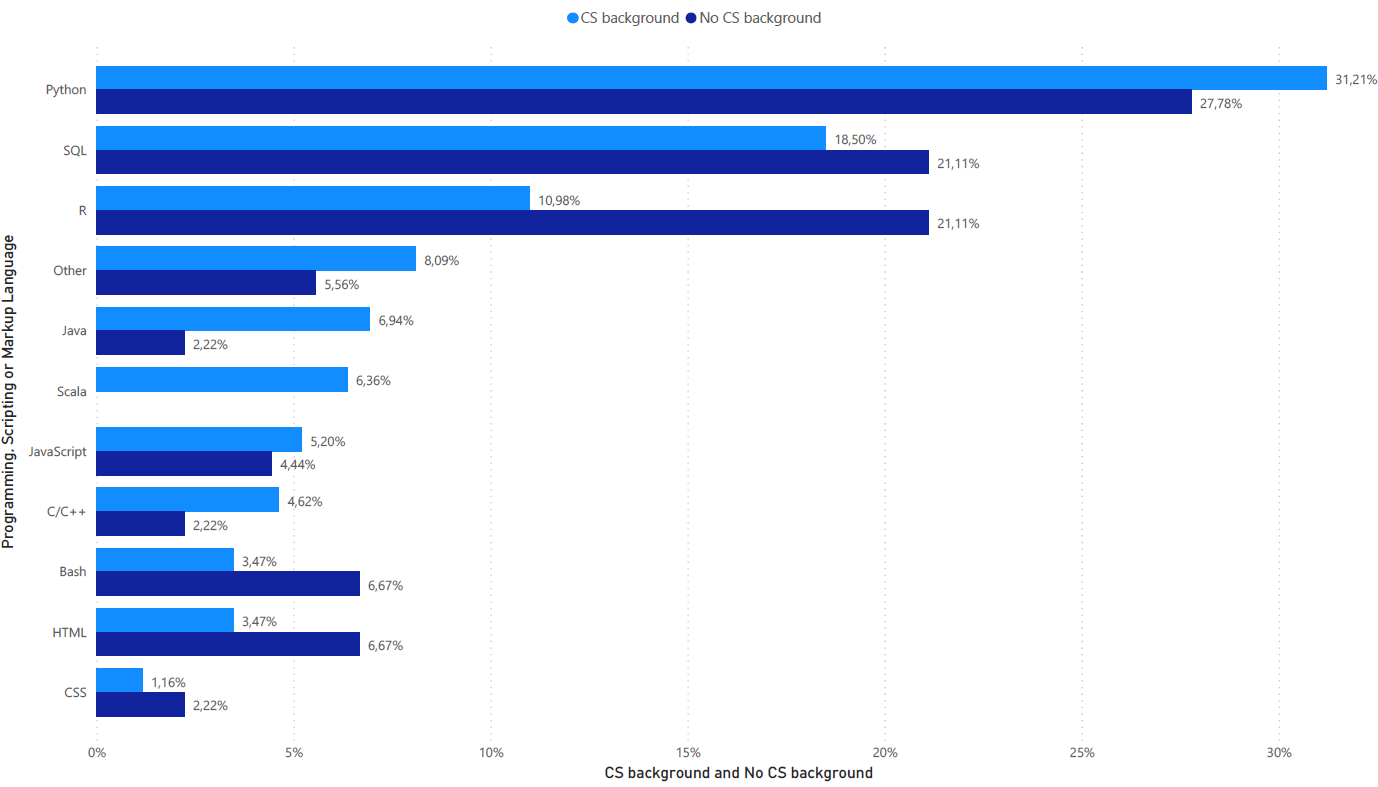}
        \caption{Programming, Scripting or Markup Language by background.}
        \label{fig:tech2}
    \end{figure}

\cref{fig:tech2} shows the choices of the participants regarding \textit{Programming, Scripting or Markup Language}. As shown, the three programming languages most used by people with a CS background and by people without a CS background coincide: \textit{Python}, \textit{SQL}, and \textit{R}. Percentage-wise, the biggest difference concerns the \textit{R} language, which was indicated by 10.98\% of the people with a CS background and by 21.11\% of the participants without a CS background. \textit{SQL} was the second most indicated language by the survey participants, which is one of the most used programming languages  for storing, manipulating, and retrieving data  in relational databases. In addition to \textit{Scala}, there were six more programming languages that were only mentioned by people with a CS background (\textit{C\#, Go, DAX, Julia, USQL,} and \textit{Visual Basics}), and two programming languages that were only mentioned by people without CS background (\textit{SAS} and \textit{Spark}). Given this information, we can conclude that \textbf{Python and SQL are the two most used programming languages by data science professionals and that people with a CS background use a greater diversity of programming languages than people without a background in CS}.

\paragraph{Machine Learning Frameworks/Libraries/Tools}

\begin{figure}[!tbhp]
        \centering
        \includegraphics[width=1\columnwidth]{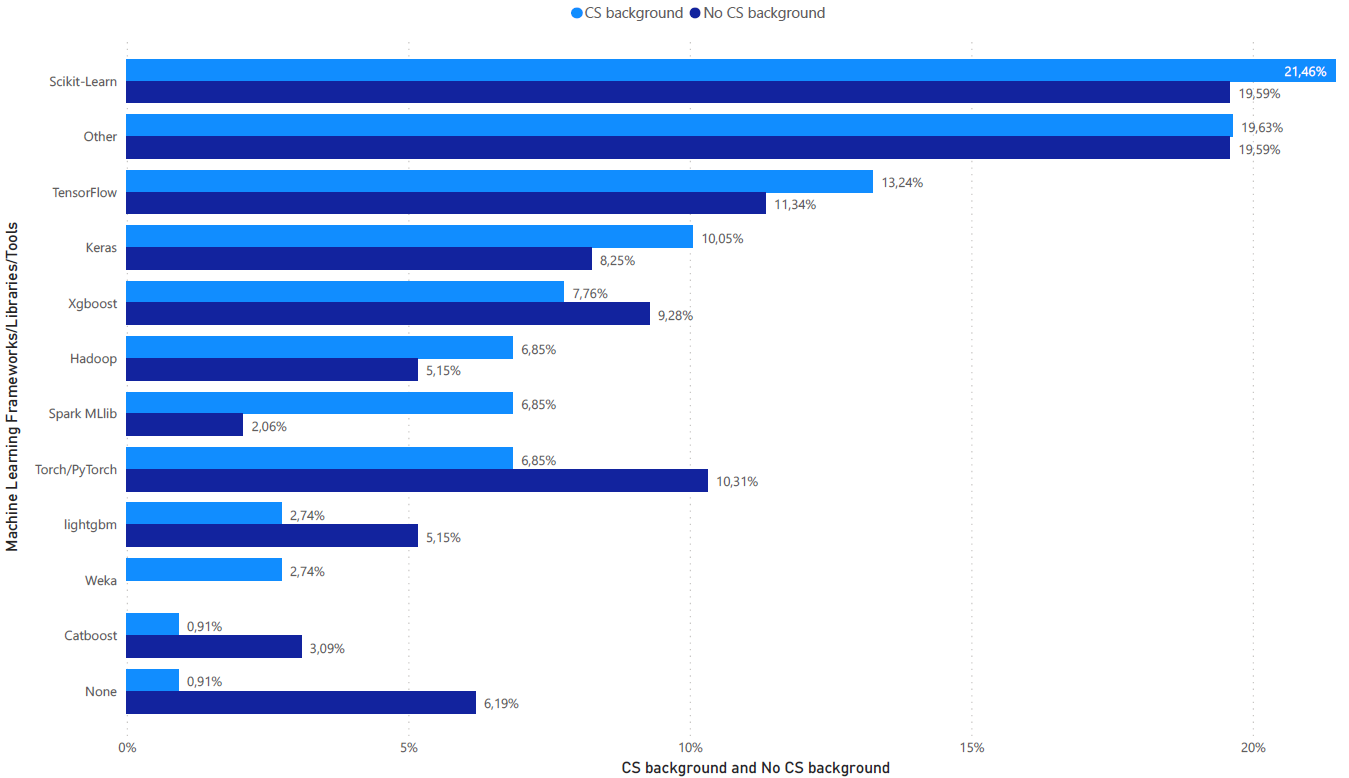}
        \caption{Machine Learning Frameworks/Libraries/Tools by background.}
        \label{fig:tech3}
    \end{figure}

\cref{fig:tech3} shows the participants' choices regarding \textit{Machine Learning Frameworks/Libraries/Tools}. As can be seen, concerning the group of participants with a CS background, the most indicated options were the library \textit{scikit-learn} (21,46\%), the open-source platform \textit{Tensorflow} (13,24\%), and the artificial neural networks library \textit{Keras} (10,05\%). In the group of people without CS background, \textit{scikit-learn} (19,59\%) and \textit{Tensorflow} (11,34\%) appear again as the first and second most indicated options, and in third place appears the machine learning framework \textit{Torch/PyTorch} (10,31\%). All four share the fact that they allow the application of machine learning techniques using \textit{Python} as a base, which once again reinforces the preference of data science professionals for \textit{Python}. Also in this section, some options were only mentioned by participants in one of the groups, with 10 being mentioned only by people with a CS background, and 3 being mentioned only by people without a CS background. Finally, it is important to note that the percentage of participants without a CS background who indicated not using any type of \textit{Frameworks/Libraries/Tools} is 6.19\%, while in the group of people with a CS background, this percentage is only 0.91\%.

\paragraph{Statistics Packages/Tools}
    
\begin{figure}[!tbhp]
        \centering
        \includegraphics[width=1\columnwidth]{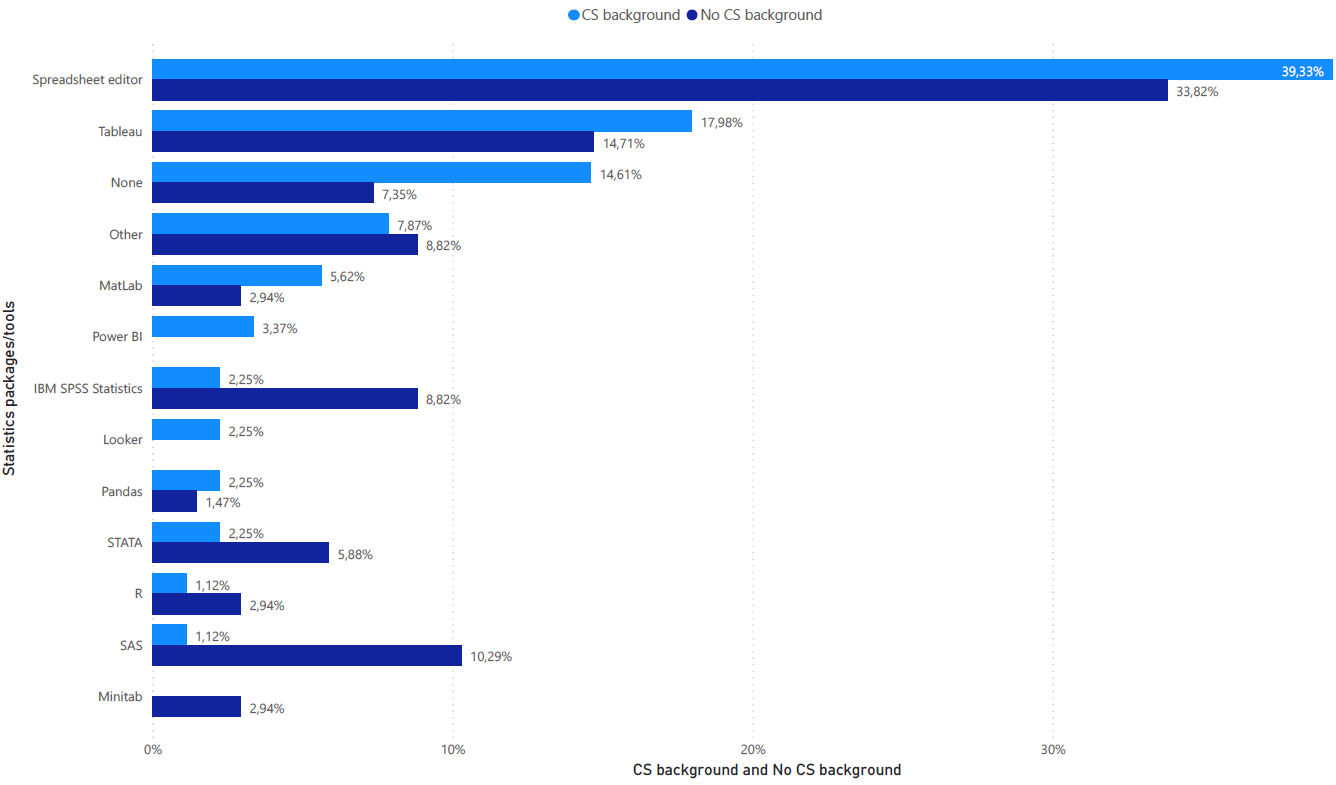}
        \caption{Statistics packages/tools by background.}
        \label{fig:tech4}
    \end{figure}

\cref{fig:tech4} shows the choices of the participants regarding \textit{Statistics Packages/Tools}. It is clear that \textit{Spreadsheet editors} are the preferred tools for statistical analysis amongst both data science professionals with CS background (39,33\%) and without CS background (33,82\%). As can be seen, \textit{Tableau}, a software that allows for visual analysis of data, is also widely used by these professionals. On the other hand, \textit{SPSS} and \textit{SAS}, despite sharing several features offered by \textit{Tableau}, are mainly used only by professionals without a CS background. Finally, 14,61\% and 7,35\% of professionals with and without CS background, respectively, indicated not using any \textit{statistics packages/tools} during their work, which amounts to 11,46\% of the respondents.

\paragraph{Data visualization Libraries/Tools}

\begin{figure}[!htbp]
        \centering
        \includegraphics[width=1\columnwidth]{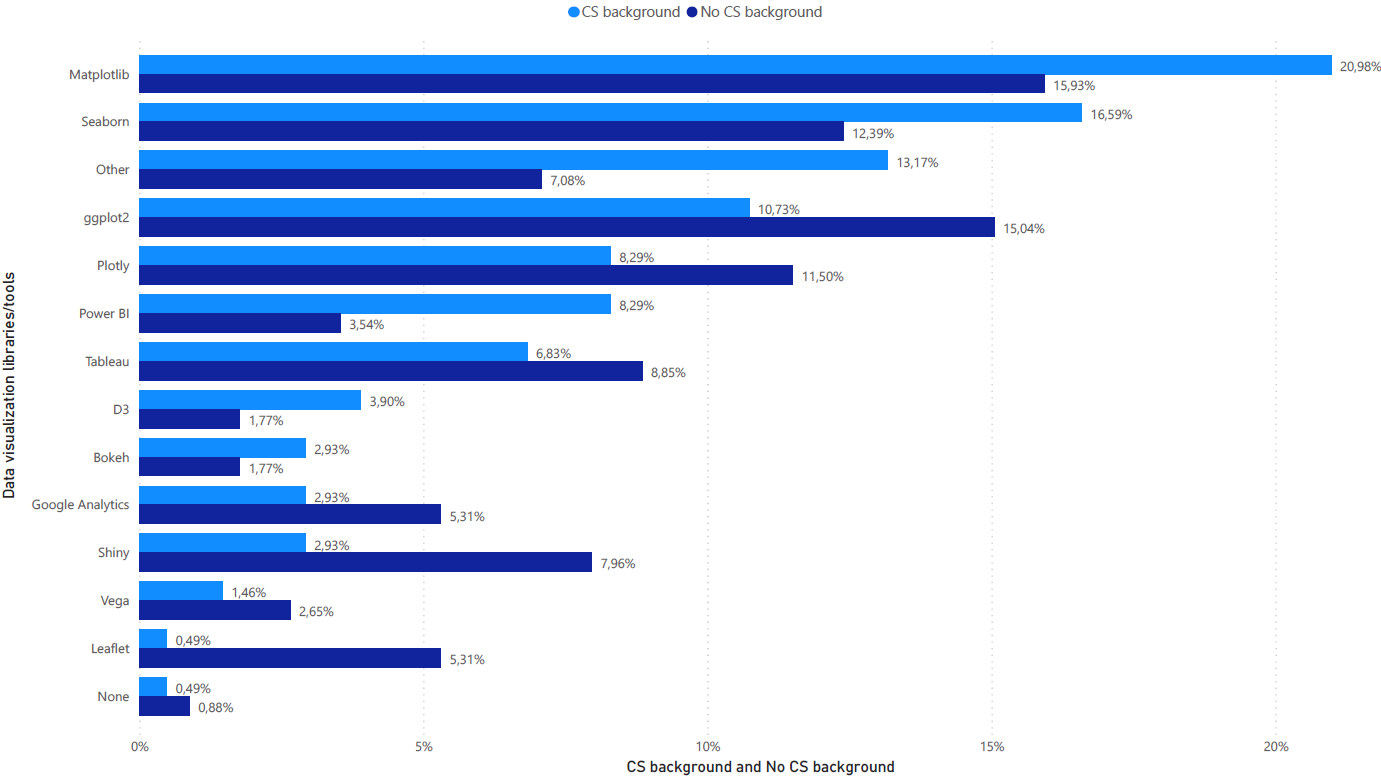}
        \caption{Data visualization libraries/tools by background.}
        \label{fig:tech5}
    \end{figure}

\cref{fig:tech5} shows the choices of the participants regarding \textit{Data visualization libraries/tools}. Once again, there is a clear predominance of \textit{Python-based} libraries being used for data visualization. The most used libraries are \textit{Matplotlib, Seaborn} and \textit{ggplot2}. Besides these, tools that provide an interactive way to create and manage visualizations, such as \textit{Power BI} or \textit{Tableau}, are also widely used. With this information, we can also infer that \textbf{almost no data science professional can conduct their work without a visualization tool}.

\section{Key Results}

In this section, we answer the RQs defined in \cref{sec:introduction}.


\paragraph{$\mathcal{RQ}_1$ What is the profile of a data science professional?}

Regarding the profile of data science professionals, we conclude that they are mostly men aged between 26 and 45 years and that there are several indicators that point to the existence of a gender gap in the area of data science.

The data also revealed that data professionals are highly qualified and most have an academic background in computer science. Besides formal education, data science professionals use online resources such as courses, blogs about data science, and online lectures from ivy league universities, to expand their knowledge and skills related to data science.

Finally, most data science professionals work as data scientists, data analysts, or on machine learning, and have up to 4 years of professional experience. They are satisfied, or very satisfied, with the jobs they do, but factors such as years of experience, academic background, and gender can influence their satisfaction.

\paragraph{$\mathcal{RQ}_2$ How does the profile of the data science professional impact their work?}

Considering the profile outlined above, we conclude that there is no characteristic that is particularly decisive with regard to the way data science professionals perform their jobs.

Data scientists spend up to half of their time programming and are generally confident in their abilities to complete all of their tasks. The tasks in which they have less experience are those that require the application of analytical methods, particularly deep learning methods. This is especially true for professionals without a computer science background. 

Finally, the situations that most affect the productivity of data scientists are the access to quality data relevant to the problems they work on, and the lack of clear questions to answer.

\paragraph{$\mathcal{RQ}_3$ Which technologies are most frequently used?}

There seems to be a slight difference in the choices between those with a CS background and those without a CS background.
Professionals with a CS background have a strong preference for Python, which is reflected in their preferred IDE as well as their choices in machine learning and data visualization technologies. Aside from Python, the R language appears to be a common choice among professionals without a computer science background.

Finally, concerning technologies used in data science, there seems to be a consensus among professionals in choosing spreadsheet editors for the statistical analysis of data.
    
    \section{Threats to validity}

In this section, we discuss threats that may affect the validity of our findings.


\paragraph{Sampling errors} These happen when the respondents of a survey over or under-represent the population of interest. If this occurs, the results obtained cannot be generalized, which poses a threat to the reliability of the findings.
    To avoid this situation, we distributed the survey on multiple platforms and the responses collected came from data science professionals from different countries, companies, and professional paths. 
    
\paragraph{Survey scope errors} These occur when the survey does not include important questions that cover all the major aspects of the study. If this occurs, it is very difficult to answer the proposed research questions.
    To avoid this, we designed our survey with help from a person with experience in data science. Moreover, before distributing the survey, we reviewed it to ensure that every question was relevant to at least one RQ. By reviewing the questions, we also confirmed that the survey was not too long, which could lead to respondents abandoning it and not submitting their answers. 
    
\paragraph{Non-response errors} These occur when the respondents of the survey fail to answer multiple and important questions, which can be accidental or intentional. Similar to the survey scope error, this can lead to losing important information that would allow us to answer the research questions.
    To avoid this, we made sure that the questions in the survey were divided into smaller sections so that the process of answering the survey did not become tedious, and that the questions were mandatory.
    
\paragraph{Analysis and interpretation bias errors} These occur when  researchers have a personal bias in favor of a particular hypothesis, which  may lead them to manipulate the data to support the hypothesis they believe to be true.
    To avoid this and to ensure that other people are able to examine this work and came to the same conclusions, before documenting the findings, we explored different interpretations of the data and reviewed the results with outside peers that provided affirmation that our conclusions are reasonable.

\paragraph{Response bias} Participants may fill influenced to answer questions in a certain direction, which may cause the conclusions to be based on not entirely valid data.
However, with our work, we just tried to better know data scientists and were not seeking any particular result. Thus, we believe our questions did not influence particular answers and thus the self-assessment provided by the participants is reliable.

    \section{Conclusions and future work}
    Our main goal was to get to know data science professionals better and to understand if their academic background has an impact on the way they perform their jobs and on the technologies they use.

To do so, we conducted an online survey containing questions on diverse topics such as academic background, professional situation, self-assessment on data science-related tasks, difficulties faced during work, and technologies used. To avoid sampling issues that could reduce the generalizability of our findings, our survey was distributed online which made it possible to collect responses from data science professionals around the world.

The obtained knowledge allowed us to trace the profile of data science professionals and to conclude that people in data science are generally highly qualified, and although most of them have a computer science academic background, their academic past has little impact on the way they work. We found that the most common difficulties are shared among all professionals, namely the access to quality data relevant to the problems they work on, and the application of deep learning techniques. We also discovered evidence of a gender gap in data science, as the number of female professionals is much lower than the number of male ones, and job satisfaction is also lower among female professionals.

In the future, we hope to use this knowledge to propose new work methodologies and tools that can be productively used by people performing data science tasks within a visual, familiar, and user-friendly environment.

\section*{Acknowledgments}

This work is financed by National Funds through the Portuguese funding agency, FCT - Fundação para a Ciência e a Tecnologia, within project LA/P/0063/2020.

This research was supported by the doctoral Grant SFRH/BD/151366/2021 financed by the Portuguese Foundation for Science and Technology (FCT), and with funds from Portugal 2020, under MIT Portugal Program.

\bibliographystyle{plain}
\bibliography{main.bib}

\begin{thebibliography}{10}

\bibitem{Shane2009}
Shane Allua and Cheryl~Bagley Thompson.
\newblock {Inferential Statistics}.
\newblock {\em Air Medical Journal}, 28(4):168--171, 7 2009.

\bibitem{Bickman2008}
Leonard Bickman and Debra~J. Rog.
\newblock {\em {The SAGE Handbook of Applied Social Research Methods}}.
\newblock SAGE Publications, Inc, 2 edition, 2008.

\bibitem{Cao2017}
Longbing Cao.
\newblock Data science: A comprehensive overview.
\newblock {\em ACM Comput. Surv.}, 50(3), jun 2017.

\bibitem{Chu2016}
Xu~Chu, Ihab~F. Ilyas, Sanjay Krishnan, and Jiannan Wang.
\newblock {Data cleaning: Overview and emerging challenges}.
\newblock In {\em Proceedings of the ACM SIGMOD International Conference on
  Management of Data}, pages 2201--2206. Association for Computing Machinery, 6
  2016.

\bibitem{Davenport2012}
Thomas~H. Davenport and D.~J. Patil.
\newblock {Data scientist: The sexiest job of the 21st century}.
\newblock {\em Harvard Business Review}, 90(10):5, 2012.

\bibitem{Dhar2014}
Vasant Dhar, Matthias Jarke, and Jürgen Laartz.
\newblock {Big Data}.
\newblock {\em Business {\&} Information Systems Engineering}, 6(5):257--259,
  10 2014.

\bibitem{Fisher2009}
Murray~J. Fisher and Andrea~P. Marshall.
\newblock {Understanding descriptive statistics}.
\newblock {\em Australian Critical Care}, 22(2):93--97, 5 2009.

\bibitem{Grupe1995}
Fritz~H. Grupe and M.~Mehdi Owrang.
\newblock {Data base mining: Discovering new knowledge and competitive
  advantage}.
\newblock {\em Information Systems Management}, 12(4):26--31, 1995.

\bibitem{Harris2013}
Harlan Harris, Sean Murphy, and Marck Vaisman.
\newblock {\em {Analyzing the Analyzers: An Introspective Survey of Data
  Scientists and Their Work}}.
\newblock O'Reilly Media, Inc., 1 edition, 2013.

\bibitem{Holst2019}
Arne Holst.
\newblock {Data created worldwide 2010-2025 | Statista}, 2019.

\bibitem{Kandel2011}
Sean Kandel, Jeffrey Heer, Catherine Plaisant, Jessie Kennedy, Frank van Ham,
  Nathalie~Henry Riche, Chris Weaver, Bongshin Lee, Dominique Brodbeck, and
  Paolo Buono.
\newblock {Research directions in data wrangling: Visualizations and
  transformations for usable and credible data}.
\newblock {\em Information Visualization}, 10(4):271--288, 2011.

\bibitem{Kim2017}
Hae-Young Kim.
\newblock {Statistical notes for clinical researchers: Chi-squared test and
  Fisher's exact test}.
\newblock {\em Restorative Dentistry {\&} Endodontics}, 42(2):152, 3 2017.

\bibitem{Kim2016}
Miryung Kim, Thomas Zimmermann, Robert DeLine, and Andrew Begel.
\newblock {The emerging role of data scientists on software development teams}.
\newblock {\em Proceedings - International Conference on Software Engineering},
  14-22-May-:96--107, 2016.

\bibitem{Kitchenham2003}
Barbara Kitchenham and Shari~Lawrence Pfleeger.
\newblock {Principles of survey research part 6}.
\newblock {\em ACM SIGSOFT Software Engineering Notes}, 28(2):24, 2003.

\bibitem{Kubina2015}
Milan Kubina, Michal Varmus, and Irena Kubinova.
\newblock {Use of Big Data for Competitive Advantage of Company}.
\newblock {\em Procedia Economics and Finance}, 26:561--565, 2015.

\bibitem{Mack2015}
Leonard Mack and David Tarrant.
\newblock {D1.4 Study Evaluation Report 2}.
\newblock Technical report, European Commission, 2015.

\bibitem{Muller}
Heiko M{\"u}ller and Johann~Christoph Freytag.
\newblock {\em Problems, methods, and challenges in comprehensive data
  cleansing}.
\newblock Professoren des Inst. F{\"u}r Informatik, 2005.

\bibitem{Muller2019}
Michael Muller, Ingrid Lange, Dakuo Wang, David Piorkowski, Jason Tsay, Q.~Vera
  Liao, Casey Dugan, and Thomas Erickson.
\newblock How data science workers work with data: Discovery, capture,
  curation, design, creation.
\newblock In {\em Proceedings of the 2019 CHI Conference on Human Factors in
  Computing Systems}, CHI '19, page 1–15, New York, NY, USA, 2019.
  Association for Computing Machinery.

\bibitem{Parkins2017}
David Parkins.
\newblock {Regulating the internet giants: The world's most valuable resource
  is no longer oil, but data. The Economist (United Kingdom)}, 2017.

\bibitem{Perry2019}
Tekla~S. Perry.
\newblock {Demand and Salaries for Data Scientists Continue to Climb}, 2019.

\bibitem{Provost2013}
Foster Provost and Tom Fawcett.
\newblock {Data Science and its Relationship to Big Data and Data-Driven
  Decision Making}.
\newblock {\em Big Data}, 1(1):51--59, 3 2013.

\bibitem{Rahm2014}
Erhard Rahm and Hong~Hai Do.
\newblock {Data Cleaning: Problems and Current Approaches Erhard}.
\newblock {\em IEEE Transactions on Cloud Computing}, 2(1):1--1, 2014.

\bibitem{Rexer2015}
Karl Rexer, Paul Gearan, and Heather Allen.
\newblock {2015 Data Science Survey}.
\newblock Technical report, Rexer Analytics, 6 2015.

\bibitem{Schroeck2012}
Michael Schroeck, Rebecca Shockley, Janet Smart, Dolores Romero-Morales, and
  Peter Tufano.
\newblock {Analytics: The real-world use of big data}.
\newblock Technical report, IBM Institute for Business Value, 2012.

\bibitem{VanDenBroeck2005}
Jan Van Den~Broeck, Solveig~Argeseanu Cunningham, Roger Eeckels, and Kobus
  Herbst.
\newblock {Data cleaning: Detecting, diagnosing, and editing data
  abnormalities}, 2005.

\bibitem{Williams2018}
Ayshea Williams.
\newblock {How Big Data is Creating Competitive Advantage | K2 Partnering
  Solutions}, 2018.

\bibitem{Wohlin2012}
Claes Wohlin, Per Runeson, Martin H{\"{o}}st, Magnus~C. Ohlsson, Björn
  Regnell, and Anders Wessl{\'{e}}n.
\newblock {\em {Experimentation in Software Engineering}}.
\newblock Springer Berlin Heidelberg, Berlin, Heidelberg, 1 edition, 2012.

\bibitem{Wongsuphasawat2019}
Kanit Wongsuphasawat, Yang Liu, and Jeffrey Heer.
\newblock Goals, process, and challenges of exploratory data analysis: An
  interview study.
\newblock {\em CoRR}, abs/1911.00568, 2019.

\bibitem{Maire2016}
Oliver Wyman.
\newblock {The Data Science Revolution That's Transforming Aviation}, 2017.

\bibitem{Zhang2020}
Amy~X. Zhang, Michael~J. Muller, and Dakuo Wang.
\newblock How do data science workers collaborate? roles, workflows, and tools.
\newblock {\em CoRR}, abs/2001.06684, 2020.

\end{thebibliography}

\end{document}